\documentclass[11pt]{article}
\usepackage[english]{babel}
\usepackage[utf8]{inputenc}
\usepackage[T1]{fontenc}
\usepackage{authblk}
\usepackage{graphicx}
\usepackage{amsmath,mathrsfs,dsfont}
\usepackage{amsthm}
\usepackage[all]{xy}
\usepackage{graphics}
\usepackage{tensor}
\usepackage{verbatim}
\usepackage{nicefrac}
\usepackage{amsfonts}
\usepackage{amssymb,latexsym}
\usepackage{color}
\usepackage{mathrsfs,dsfont}
\usepackage{appendix}
\usepackage{slashed}
\usepackage{cite}
\usepackage{url}
\usepackage{hyperref}
\usepackage{comment}
\newcommand{\ba}{\begin{eqnarray}}
\newcommand{\ea}{\end{eqnarray}}

\begin{document}

\title{On temperature of a diamond}
\author[1]{Ravi Mistry\thanks{ravi.mistry.r@gmail.com}}
\author[1,2]{Aleksandr Pinzul\thanks{apinzul@unb.br}}
\affil[1]{Universidade de Bras\'{\i}lia, Instituto de F\'{\i}sica 70910-900, Bras\'{\i}lia, DF, Brasil}
\affil[2]{International Center of Physics C.P. 04667, Brasilia, DF, Brazil}
%\date{}
\maketitle

\begin{abstract}
We revisit the definition of the temperature of a causal diamond for the case of a free massless scalar field. The stress is given to the intrinsic, direction-dependent character of this definition. Some important limits are also discussed. 
\end{abstract}
\section{\label{sec:level1}Introduction}

Tomita-Takesaki (TT) modular theory \cite{Takesaki:1970aki} has proven to be a very important result that has found its numerous applications not only in pure mathematics but also in theoretical and mathematical physics, especially in the algebraic approach to local quantum field theory, see \cite{Borchers:2000pv,Witten:2018zxz} for an extensive discussion and applications. The physical importance of TT theory comes, roughly speaking, from the fact that, for a given state, this theory provides a natural intrinsic dynamics for the local algebras of observables. One of the most important results in this area is the explicit construction of TT theory for the algebra of observables localized in a wedge region in the Minkowski vacuum state \cite{Bisognano:1975ih,Bisognano:1976za}. In \cite{Sewell:1980}, the result of Bisognano and Wichmann was applied to give a purely algebraic derivation of the Unruh effect. (See also, \cite{Buchholz:2012wr,Buchholz:2014jta} for another recent approach to Unruh effect within the Algebraic QFT.) Later, in \cite{Connes:1994hv}, the similar ideas were used to argue that time rather than being a fundamental property of a theory, is determined by the natural dynamics given by TT theory and as such depends on a state of the system. This is the so-called ``thermal time hypothesis''.

Soon, the results of \cite{Bisognano:1975ih,Bisognano:1976za} have been extended to other regions. In \cite{Buchholz:1977ze}, the modular theory for a free massless scalar field localized in the forward light cone was constructed, while in \cite{Hislop:1981uh}, the same was done for the double cone localization. In \cite{Martinetti:2002sz}, the results of \cite{Hislop:1981uh} were interpreted from the point of view of the thermal time hypothesis \cite{Connes:1994hv}, leading to the notion of the local temperature of a causal diamond (see also \cite{Martinetti:2008ja}). In \cite{Arias:2016nip}, the question of a causal diamond's thermodynamics, including its temperature, was studied, also in the framework of the modular theory, through the notion of the relative entropy.

In this letter, we revisit the problem of a diamond's temperature. The main goal is to clarify some points in the setting closely related to the thermal time hypothesis and compare this approach to the one in \cite{Arias:2016nip}. The main stress is given to the intrinsic, local and directional notion of the temperature. The plan of the paper is as follows. In the next section, after a quick review of the set-up and the known results, we calculate the local temperature for the family of the preferred observers in a causal diamond. Then, in section \ref{large L}, we discuss somewhat counterintuitive properties of different limits. We conclude with the discussion of the results.

\section{Local temperature}

\subsection{Preliminaries and the wedge example}
For any von Neumann algebra and any normal state on it, the Tomita-Takesaki theory provides a natural, intrinsically defined dynamics. With respect to this dynamics, the state that entered the construction becomes a KMS thermal state. There is extensive literature on the subject, as a nice introduction with various earlier applications, we could recommend a collection of the lectures delivered by the major players in the field at the International Enrico Fermi school \cite{FermiSchool}. One of the most important results in the Algebraic Quantum Field Theory (see \cite{Brunetti:2015vmh} for the introduction and some recent developments) was an explicit construction of the TT theory for the algebra of observables localized in a wedge region of the Minkowski space-time, $M^{1,3}$, in the vacuum state. Here by a (right) wedge, $W_R$, we mean the following space-time region
\ba\label{wedge}
W_R = \{x \in M^{1,3} : x^1 > |x^0|\}.
\ea
Then, as it was shown in \cite{Bisognano:1975ih,Bisognano:1976za}, the Tomita-Takesaki theory for the local algebra of observables, $\mathcal{R}(W_R)$, is given by the modular Hamiltonian $H = 2\pi K_1$, where $K_1$ is just the boost in the 1-direction, and the modular conjugation $J= R_1 (\pi)\Theta$, where $R_1$ is a rotation around the 1-direction and $\Theta$ is the CPT conjugation. This means that there exists a natural dynamics relative to the vacuum state
\ba\label{TT_wedge}
\forall a\in \mathcal{R}(W_R),\ \mathrm{e}^{-it K_1} a\, \mathrm{e}^{it K_1} =: a(t)\ \mathrm{is\ again\ in}\ \mathcal{R}(W_R).
\ea
Conjugation by $J$ just takes $\mathcal{R}(W_R)$ to its commutant, which is in this case $\mathcal{R}(W_L)$. There are several very important points about (\ref{TT_wedge}).

1) The Tomita-Takesaki parameter, $t$, gives the natural, intrinsic, time flow associated with TT dynamics, so, in a sense, $t$ is the ``physical'' time defined by the vacuum state (the thermal time hypothesis \cite{Connes:1994hv}).

2) With respect to this dynamics, the Minkowski vacuum state is a thermal, i.e. KMS, state with the temperature $T=\frac{1}{2\pi}$ (this specific value is somewhat artificial and depends on the normalization of the Hamiltonian in (\ref{TT_wedge})).

3) Very important is the fact that in this specific case, the TT modular flow has a geometric realization: the space-time representation of the boost $K_1$ gives the orbits that could be interpreted as the trajectories of the naturally (i.e. in agreement with the TT dynamics) moving observers. Of course, these orbits correspond to the hyperbolic uniformly accelerated motion. This geometric interpretation is very special and exists in a very limited number of cases, see e.g. \cite{Brunetti:1992zf} and references therein.

As one of the consequences of these properties, one can naturally derive the Unruh temperature of the wedge as the temperature of the thermal state as seen by the natural observers. We will give the derivation in the local form that will admit an immediate generalization to the case of the causal diamond below.

Before we proceed, we would like to clarify what we mean by "a natural observer will see that the temperature is...". As we already indicated, by a natural observer we understand an idealized observer moving along the trajectory determined by the TT dynamics. In particular, its proper time is related to the TT parameter. To measure temperature, of course, it should be coupled to the field under consideration. Though the problem of measurement in algebraic QFT has not been settled completely (as to be universally accepted), the interpretation of an accelerated observer (for example, the one moving along some TT trajectory) as a (non-ideal) local thermometer has received a rather careful treatment in \cite{Buchholz:2012wr,Buchholz:2015fqa}. In particular, it was stressed that such an observer will see the \textit{local} temperature and this should not be confused with the global temperature of a state (which, e.g., in the case of the Minkowski vacuum will remain zero). Only in this sense, our local temperature that we define below, is the Unruh temperature \cite{Unruh:1976db}. This said, of course, the more careful physical analysis should be carried out. In this work, we will restrict ourselves to a more formal mathematical analysis.

We now return to take one more look at the wedge. Under the action of $V(t)=\mathrm{e}^{-it K_1}$, an arbitrary point of the wedge, $(x^0,x^1,x^2,x^3)$, will move along the orbit
\ba\label{trajectory_wedge}
(x^0 (t),x^1 (t),x^2 (t),x^3 (t)) = (x^0 \cosh t + x^1 \sinh t,x^1 \cosh t + x^0 \sinh t,x^2,x^3),
\ea
which is just a hyperbola in $(x^2,x^3)$ constant plane, $-(x^0 (t))^2+(x^1 (t))^2=w^2 = \mathrm{const}$, with $w=\frac{1}{a}$ being the inverse constant acceleration. The infinitesimal translation in proper time, $\tau$, then is given by
\ba\label{proper time_wedge}
\mathrm{d}\tau^2 = (-(x^0)^2+(x^1)^2) \mathrm{d}t^2 \equiv w^2 \mathrm{d}t^2 .
\ea
This shows that the proper time for a natural observer differs from the ``physical'' Tomita-Takesaki time just by the choice of a scale, $\frac{\mathrm{d}\tau}{\mathrm{d}t} = w = \frac{1}{a}$. Another way to say this is that the natural Hamiltonian for an observer moving along the TT orbit is given by $2\pi a K_1$. Comparing this with the discussion above, we immediately conclude that this observer will see the Minkowski vacuum state as the KMS state at the temperature $T=\frac{a}{2\pi}$. This is, of course, the well-known result.

A couple of comments are in order. First of all, the Tomita-Takesaki theory guarantees that each (faithful normal) state is KMS with respect to some dynamics for any given algebra of observables localized in some space-time region. What is special in the case of a wedge is that this dynamics has a clear geometric interpretation leading to the TT orbits along which the preferred observers move. This will be not true in general.\cite{Borchers:2000pv,Brunetti:1992zf} Secondly, our derivation is purely local and does not depend on the actual form of these TT orbits. As long as we know that TT modular theory is given geometrically, the derivation applies. What is special in the case of a wedge is that the global TT time is proportional to proper times of the preferred observers, leading to the temperature being constant along each preferred trajectory.

Another important observation is that it would be not entirely correct to call the temperature we just calculated, a local temperature. By the way it is defined, it is more of a temperature along the preferred trajectory. In other words, some other observer passing through the same space-time point as a preferred one following along the TT trajectory will not measure the same temperature at that point. So, in a sense, it is more natural to speak about an ``inverse temperature vector field'', $\beta^\mu (x)$, which is just a field tangent to the TT trajectories and for the case of a wedge it is of the form, cf. (\ref{trajectory_wedge})
\ba
\beta^\mu (x) = (x^1 , x^0, 0, 0).
\ea
Here $x^0$ and $x^1$ are the coordinates of an \textit{arbitrary} point in the wedge (not just the ``initial data'' corresponding to $t=0$).
Then the local \textit{directional} temperature at the point $x$ will be given by
\ba\label{temperature_W}
T^{-1} = 2\pi\|\beta\| \ \Rightarrow\ T= \frac{a}{2\pi},
\ea
i.e. the usual Unruh temperature (here, $\|\beta\|=(|\beta\cdot \beta|)^{1/2}$, with the dot in $\beta\cdot \beta$ representing the usual Minkowski inner product.). This interpretation is similar to the one emerging in \cite{Arias:2016nip} from a different approach, which we explain briefly below.

\subsection{Directional local temperature of a diamond}

While the problem of constructing TT theory for a general massive field, localized in a causal diamond region, still remains open, see, e.g. \cite{Bostelmann:2022yvj}, the answer for the free massless case has been known for some time \cite{Hislop:1981uh} (see, also \cite{Longo:2020amm}, for some recent developments). In \cite{Hislop:1981uh}, it was shown that the conformal map that maps a wedge to a causal diamond (also called a double cone) can be realized as a unitary operator on the Hilbert space of a free massless scalar field. Due to this, one has a concrete spatial isomorphism between the algebra of observables localized in a wedge and the one localized in a diamond. This spatial isomorphism is trivially used to map between the corresponding modular theories. From this, it is clear that the diamond TT theory also has a geometric interpretation introducing a family of the preferred observers.

The concrete realization of what is described above is as follows. Let $W_R$ be the right wedge as defined in (\ref{wedge}) and the double cone (or the bounded diamond), $O$, be defined by
\ba
O = (V_+ - e_0)\cap (V_- + e_0),
\ea
where $V_{\pm}$ are the forward/backward lightcones and, in general, $e_\mu$ is a unit vector in $\mu$-direction.
Define a conformal transformation (a ray inversion), $\rho$,
\ba
\rho (x) := \left(-\frac{x^0}{x\cdot x}, -\frac{\vec{x}}{x\cdot x}\right).
\ea
Then the map from $W_R$ to $O$ is given by
\ba\label{mapWO}
O = \rho\left(W_R + \frac{1}{2}e_1\right) - e_1 .
\ea
While translation in $1$-direction is guaranteed (by the axioms) to be represented by a unitary operator, $T_1$, on the corresponding Hilbert space, the same is not true for $\rho$. The main result of \cite{Hislop:1981uh} is that for the case of a free massless field, there exists a unitary operator, $U_\rho$, providing a spatial map between the algebras of observables, $V = T_1 \left(-1\right) U_{\rho}\, T_1 \left(\frac{1}{2}\right)$. With the help of this unitary operator, we can map the TT of the wedge (\ref{TT_wedge}) to the one of the diamond:
\ba\label{TT_diamond}
V_O (t) = V V_{W_R} (t) V^{-1} = T_1 \left(-1\right) U_{\rho}\, T_1 \left(\frac{1}{2}\right) \mathrm{e}^{-it K_1} T_1 \left(-\frac{1}{2}\right) U_{\rho}\, T_1 \left(1\right) = \exp \left( i \frac{t}{2}(S_0 - P_0)\right),
\ea
where $S_\mu$ and $P_\mu$ are the generators of the special conformal transformation and translations, respectively. Now, using (\ref{mapWO}), we can map the TT trajectories in the wedge (\ref{trajectory_wedge}) to the corresponding ones in the diamond, with the result for the TT trajectory passing through the point $z_{\pm} = x^0 \pm r$ at $t=0$ given by \cite{Hislop:1981uh,Longo:2020amm}
\ba\label{trajectory_diamond}
z_{\pm} (t):= x^0 (t) \pm r (t) = \frac{z_{\pm}\cosh \frac{t}{2} + \sinh \frac{t}{2}}{z_{\pm}\sinh \frac{t}{2} + \cosh \frac{t}{2}}.
\ea
(Here, the fact that the action of a map (\ref{mapWO}) on the angular variables is trivial was used.)

Now, when we have the explicit expression for the TT trajectory (\ref{trajectory_diamond}), we can repeat almost verbatim the analysis of the wedge case performed in the previous subsection. After some trivial algebra, the infinitesimal change in proper time along the TT trajectory is given by
\ba
\mathrm{d}\tau^2 = \frac{1}{4}\left(1 - z_+ (t)^2\right)\left(1 - z_- (t)^2\right)\mathrm{d}t^2 .
\ea
The interpretation is completely analogous to the one in the wedge case: from
\ba
\frac{\mathrm{d}\tau}{\mathrm{d}t} = \frac{1}{2}\sqrt{\left(1 - z_+ (t)^2\right)\left(1 - z_- (t)^2\right)},
\ea
we conclude that a preferred observer moving along a TT trajectory, at the point $z_\pm (t)$, will measure the local temperature (see the discussion before (\ref{trajectory_wedge}) and in \cite{Buchholz:2012wr,Buchholz:2015fqa})
\ba\label{temperature_xt}
T(x(t)) = \frac{1}{\pi \sqrt{\left(1 - z_+ (t)^2\right)\left(1 - z_- (t)^2\right)}}.
\ea
As before, this result can be conveniently expressed in terms of the ``inverse temperature vector field'', $\beta^\mu (x)$, which in this case takes the form (keeping just the non-trivial $\pm$-components)
\ba
\beta_{\pm}(x(t)) = \frac{1-z_{\pm}^2}{2(z_{\pm}\sinh \frac{t}{2} + \cosh \frac{t}{2})^2} \equiv \frac{1}{2}\left(1-z_{\pm}(t)^2 \right).
\ea
Because the initial point of each TT trajectory, $z_\pm$, along with the global modular parameter, $t$, (along with the remaining angular coordinates) uniquely parameterizes the diamond, we can formulate the result as follows:

The local directional temperature measured by a preferred TT observer passing through the point $z_\pm$ is given at that point by
\ba\label{temperature_x}
T(x) = \frac{1}{2\pi\|\beta\|} \equiv \frac{1}{\pi \sqrt{\left(1 - z_+ ^2\right)\left(1 - z_- ^2\right)}},
\ea
where the ``inverse temperature vector field'', $\beta_\pm (x)= \frac{1}{2}\left(1-z_{\pm}^2\right)$.

The result (\ref{temperature_xt},\ref{temperature_x}) should be compared to the one obtained in \cite{Martinetti:2002sz}: it is essentially the same result, but our derivation is purely intrinsic. In \cite{Martinetti:2002sz}, the same result is obtained by comparing the TT trajectories with the trajectories of globally uniformly accelerated observers. As such, it heavily relies on the fact that (\ref{trajectory_diamond}) are also describing the hyperbolic motion, but now parameterized by the TT time for the diamond, $t$. We want to stress that as soon as the TT, or rather its geometric action, is given, there is no need in introducing another set of external observers. The directional temperature associated with the family of the preferred TT observers can be introduced intrinsically. We will have more to say on this in the next section.

Our definition of the directional temperature is very similar to the one appearing in \cite{Arias:2016nip}. In that work, the local temperature of a general region is defined, essentially, by the laws of thermodynamics, where the entropy is taken to be the relative entropy between the Minkowski vacuum and its local perturbation, but the natural dynamics of the region is still due to the Tomita-Takesaki modular theory. Then, in the case when a geometric TT action is given, the result for the relative entropy, $S(pert|vac)$, turns out to be
\ba
S(pert|vac) = 2\pi P_\mu \beta^\mu ,
\ea
where $P_\mu$ is 4-momentum and $\beta^\mu$ is the vector field tangent to the TT trajectories. We see that this leads to our results (\ref{temperature_W}) and (\ref{temperature_xt}) (or (\ref{temperature_x})) for the wedge and diamond cases, respectively, if one interprets $\beta^\mu$ as ``inverse temperature vector field''.

\section{Large \emph{L} limits} \label{large L}

Naively, one could expect that if the size of a causal diamond is taken to be very large, the effect of the finite size should be less and less important. That this limit is much trickier could be seen from the following consideration: if one just takes the limit of a diamond to infinity, the expected theory should look like the one in Minkowski space-time. If, on the other hand, along with sending the size to infinity we translate a diamond in some controlled manner (as to keep a part of the boundary at a finite distance), the situation should start resembling the one of a wedge. It is the aim of this section to study such limits as well as to clarify the picture as seen from the point of view of the global TT time versus the local proper time of a preferred observer.

\begin{figure}[t]
\includegraphics[scale=0.7]{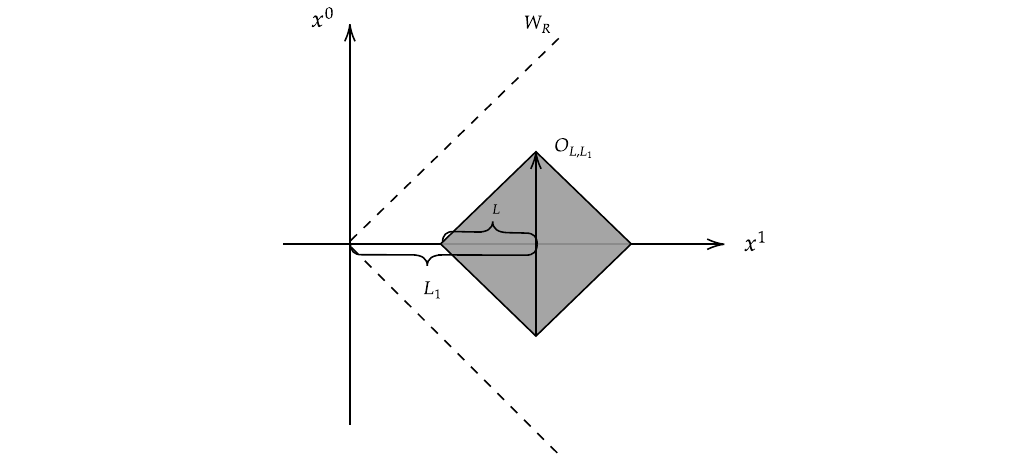}
\centering
\caption{A causal diamond of the size $L$, translated by the distance $L_1$ in the $1$-direction.}
\label{Fig1}
\end{figure}

To make the above remarks more explicit, let us consider the Tomita-Takesaki operator (\ref{TT_diamond}) for a causal diamond of an arbitrary size, $L$, ($L=1$ in (\ref{TT_diamond})) translated by a distance $L_1$ along the $1$-direction, ${O_{L,L_1}}$. While the dependence on the size of a diamond is easily obtained by the dimensional analysis, the translated TT is arrived at after some trivial, even though slightly lengthy, algebra with the result given by
\ba\label{TT_diamond_L}
V_{O_{L,L_1}} (t) = \mathrm{e}^{i L_1 P_1} V_{O} (t) \mathrm{e}^{-i L_1 P_1} = \exp \left( \frac{it}{2L} (S_0 + (L_1^2 - L^2) P_0 - 2 L_1 K_1)\right),
\ea
where all the notations should be clear from the Fig.\ref{Fig1}.

Let us consider two formal (i.e. not specifying the topology) limits of (\ref{TT_diamond_L}). First, consider the limit of a large diamond at a fixed position. This corresponds to $L\gg 1$, while $L_1$ is kept finite and $L\gg L_1$. Then the naive limit of (\ref{TT_diamond_L}) has the form
\ba\label{TT_diamond_L_1}
V_{O_{L,L_1}} (t) \rightarrow \exp \left( -\frac{it}{2} L P_0 \right).
\ea
This should be interpreted as that $V_{O_{L,L_1}}$ looks more and more like the TT for the Minkowski space-time, i.e. the natural dynamics is given by $P_0$ while the temperature goes to zero as $1/L$ (as it should be because the Minkowski vacuum is a pure state).

The second limit corresponds to sending both, $L$ and $L_1$ to infinity in such a way that $\frac{L_1}{L}\rightarrow 1$ (It will be convenient to define $L_1^2 = L^2 + w^2$ for some $w = const$). For this case, (\ref{TT_diamond_L}) tends to
\ba\label{TT_diamond_L_2}
V_{O_{L,L_1}} (t) \rightarrow \exp \left( -{it} K_1 \right).
\ea
This should be compared to (\ref{TT_wedge}). One can see that this limit naively corresponds to the TT behaving more and more as the modular theory for a wedge.

Actually, the naive limits (\ref{TT_diamond_L_1}) and (\ref{TT_diamond_L_2}) are very deceptive. They are valid only in very tiny (relative to $L$) regions of $O_{L,L_1}$. To see this, let us instead of looking at the Tomita-Takesaki operator (\ref{TT_diamond_L}) do a similar analysis for the TT trajectories, which are given by a straightforward generalization of (\ref{trajectory_diamond})\footnote{We restrict ourselves to the TT trajectories in the $x^0-x^1$ plane. The explicit form for a more general trajectory will be more involved after a translation, but because the choice of the $1$-direction is arbitrary, we do not loose any generality of the results.}
\ba\label{trajectory_diamond_L}
\frac{z_{\pm} (t) \mp L_1}{L} = \frac{\frac{z_{\pm}\mp L_1}{L}\cosh \frac{t}{2} + \sinh \frac{t}{2}}{\frac{z_{\pm}\mp L_1}{L}\sinh \frac{t}{2} + \cosh \frac{t}{2}}\ \Rightarrow\ z_{\pm} (t)= \frac{L z_{\pm}\cosh \frac{t}{2} + (L^2 \pm L_1 z_{\pm} -L_1^2 )\sinh \frac{t}{2}}{(z_{\pm}\mp L_1)\sinh \frac{t}{2} + L\cosh \frac{t}{2} }.
\ea
Namely, we want to see for what parts of $O_{L,L_1}$ (and for what values of the TT parameter, $t$) the TT flow generated by the full modular operator (\ref{TT_diamond_L}) ``looks like'' the one generated by the corresponding limiting modular operators, (\ref{TT_diamond_L_1}) and (\ref{TT_diamond_L_2}). This is easily done starting with (\ref{trajectory_diamond_L}).

For the case of (\ref{TT_diamond_L_1}), let us take $L_1 = 0$ and $L\gg 1$. Without loss of generality, we can take $z_{\pm} = \pm r$, i.e. $x^0(0)=0$, $x^1(0)=r$. Then
\ba\label{trajectory_diamond_L1}
\ z_{\pm} (t) \rightarrow \frac{1}{2} Lt \pm r
\ea
only for a very small range of the TT parameter, $t\ll 1$ (and a much less restrictive condition, $tr \ll L$). But (\ref{trajectory_diamond_L1}) are exactly the trajectories generated by (\ref{TT_diamond_L_1}). So, the Tomita-Takesaki theory for a diamond will look like the one for the Minkowski space only in a very slim region around the ball, $x^0=0$. Closer to the center of the ball, corresponding to $r=0$, the Minkowski-like behaviour will be slightly improving, as should be expected.

For the case of (\ref{TT_diamond_L_2}), after some straightforward algebra, we get
\ba\label{trajectory_diamond_L2}
\ z_{\pm} (t) \rightarrow \pm r \mathrm{e}^{\pm t} .
\ea
This limit is valid only if $\frac{r - L_1 +L}{L}\sim\frac{r}{L}\ll 1$, i.e. if $r$ is very close (in units of $L$) to the left corner of the diamond (the same is true, of course, for the right corner).\footnote{Recall that we are considering only the TT trajectories in $x^0-x^1$ plane. So the ``corners'' here are actually wedge-like regions for some neighbourhood in $x^2$, $x^3$ directions.} But (\ref{trajectory_diamond_L2}) is exactly (\ref{trajectory_wedge}) (for $x^0(0)=0$, $x^1(0) = r$), which are the usual TT trajectories for the wedge modular operator (\ref{TT_diamond_L_2}).

Intuitively, these results are clear: the Minkowski-like behaviour should be in the region where the effect of all the boundaries of a diamond will be less felt, i.e. closer to the center; while the wedge-like TT should be reproduced when mostly only one boundary (the left one in our analysis, see the footnote) will have a significant effect. What is somewhat surprising is that the size (relative to $L$) of the corresponding regions is extremely small and almost everywhere in the diamond its TT dynamics will be radically different from either the wedge or the Minkowski ones.

At this point, it would be instructive to notice that the TT trajectories (\ref{trajectory_diamond}) (or (\ref{trajectory_diamond_L})) are actually hyperbolas. This fact was heavily used in \cite{Martinetti:2002sz} to define the diamond temperature. As we have seen, the definition of the local (directional) temperature can be given in the form that does not depend on the global form of the trajectories. Still, because the TT trajectories for a wedge, (\ref{trajectory_wedge}), also describe hyperbolic dynamics, it would be interesting to compare these two. Physically this would correspond to having two observers following the same space-time trajectory but with respect to two different dynamics (i.e. one being parameterized by the wedge TT parameter, while the other by the diamond one). As we saw in (\ref{trajectory_diamond_L2}), the only region where these two observers will have ``close'' dynamics (= measure similar temperatures) is around the wedge-like part of a diamond.

The diamond temperature (\ref{temperature_x}) is immediately generalized for a translated diamond of an arbitrary size, $O_{L,L_1}$,
\ba\label{temperature_xL}
T_{O_{L,L_1}}(x) = \frac{L}{\pi \sqrt{\left(L^2 - (z_+ - L_1)^2\right)\left(L^2 - (z_- + L_1)^2\right)}}.
\ea
\begin{figure}[t]
\includegraphics[scale=0.7]{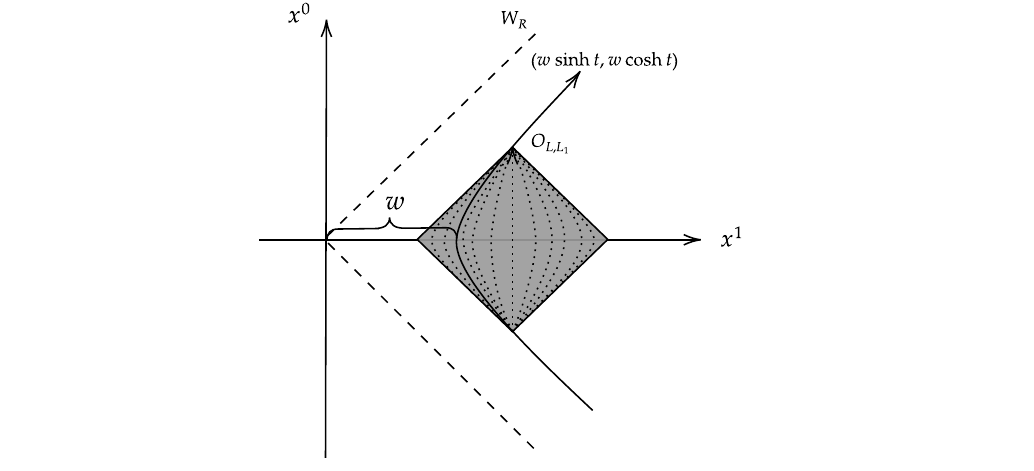}
\centering
\caption{When a diamond is translated inside a wedge by the distance $L_1 = \sqrt{L^2 + w^2}$, one and only one of the diamond TT trajectories will coincide with the standard wedge hyperbola corresponding to the motion with the constant acceleration $a=1/w$. The other diamond TT trajectories are schematically shown as dotted curves.}
\label{Fig2}
\end{figure}
For a given $L_1 = \sqrt{L^2 + w^2}$ exactly one TT trajectory of a diamond will coincide with exactly one hyperbolic trajectory of a wedge, namely with the one corresponding to the acceleration $a=1/w$, see Fig.\ref{Fig2}. Using this observation, one can compare the local directional temperatures as seen by a natural diamond and a natural wedge observers. It is more convenient to keep $L_1 = 0$ in (\ref{temperature_xL}) and just translate the wedge. Because for a given acceleration the wedge temperature is given by (\ref{temperature_W}), we need to express acceleration as a function of an arbitrary point of a diamond, as in (\ref{temperature_x}), $a=a(z_+,\,z_-)$. Let $z^w_\pm$ be the coordinates on the wedge translated by $L_1=\sqrt{L^2+w^2}$ in such a way that the uniformly accelerated observer exactly passes through the diamond point $(z_+,\,z_-)$, see Fig. \ref{Fig2}. Then the relation between $z^w_\pm$ and $z_\pm$ is trivially given by $z^w_\pm=z_\pm\pm L_1\equiv z_\pm \pm \sqrt{L^2+w^2}$. But $z_\pm^w$ satisfy the hyperbolic relation, see (\ref{trajectory_wedge}) and below:
\ba
-w^2=z^w_+ z^w_-=\left(z_+ + \sqrt{L^2+w^2}\right) \left(z_- - \sqrt{L^2+w^2}\right).
\ea
Solving this for $w^2$, we easily get
\ba\label{accelaration}
w^2(z_+,\,z_-)=\frac{(L^2-z^2_+)(L^2-z^2_-)}{(z_+ - z_-)^2}\implies a^2(z_+,\,z_-) = \frac{(z_+ - z_-)^2}{\left(L^2 - z_+^2\right)\left(L^2 - z_-^2\right)},
\ea
which leads to a very simple result
\ba\label{ratio}
\frac{T_W}{T_O} = \frac{z_+ - z_-}{2L} \equiv \frac{r(z)}{L}.
\ea
Here $r(z)$ is the radial component of an arbitrary point, $z=(z_+,\,z_-)$, of a diamond $O_{L,0}=O$. So, in the complete agreement with the discussion above, we see that $\frac{T_W}{T_O} \rightarrow 1$ only in the tiny region where $r \rightarrow L$. Note that (\ref{ratio}) does not depend on the restriction to the $x^0-x^1$ plane and it is valid for any arbitrary point of the diamond. From (\ref{trajectory_diamond}) (or (\ref{trajectory_diamond_L})) we can find $r(z)$ as a function of the TT parameter along the unique TT trajectory passing through $z$\footnote{Here the TT trajectory that passes through $z$ at some $t$, at $t=0$ passes through $x^0(0)=0$ and $r(0) = r$.}
\ba\label{r_t}
r(t) = \frac{r}{\left( 1 - \frac{r^2}{L^2} \right)\sinh^2\frac{t}{2} + 1}.
\ea
Eq.(\ref{r_t}) combined with (\ref{ratio}) shows that only when $r$ is extremely close to $L$ the ratio $\frac{T_W}{T_O}$ will be close to 1 during some extended TT time, $t$. More precisely if ($\Delta r := L - r$)
\ba
\Delta r \sinh^2\frac{t}{2}\ll L ,
\ea
the wedge and the diamond observers will agree on the local temperatures they are measuring. After that, the ratio (\ref{ratio}) will exponentially quickly go to zero.

\section{Conclusion}\label{Conclusion}

In this note, we revisited the question of the intrinsic definition of local temperatures of certain space-time regions. In the framework of the Algebraic QFT, we studied the cases of a wedge and a causal diamond. Though the results, of course, agree with the ones obtained earlier in \cite{Sewell:1980} and \cite{Martinetti:2002sz}, the main stress in our analysis is on the intrinsic and direction dependent nature of the local temperature. By intrinsic, we mean the following: once the Tomita-Takesaki theory for some state is known, one does not need any external reference observers (as the global hyperbolic ones in \cite{Martinetti:2002sz}) to define the local temperature. Unfortunately, the demonstration of this result crucially depends on the geometric realization of the TT modular flow, which is the case for a wedge and a diamond.

The directional dependence of our definition of the local temperature is obvious: the only natural way of defining such temperature could be given for a family of the preferred observers given by the geometrical action of the TT modular flow. In such a way, the thermal properties of a diamond are probed only along the trajectories of these observers, leading to the natural definition of the inverse temperature vector field, $\beta_\mu$. In this point, our consideration comes closely to the one done in \cite{Arias:2016nip}.

Physically, one of the most important (and somewhat counterintuitive) results of our study is the explicit demonstration that the natural dynamics, i.e. given by the TT modular theory, of the algebra of observers of a diamond is radically different from the dynamics of the imbedding space-time regions (a wedge or the whole Minkowski space-time). We showed that only in a tiny part of the diamond the dynamics will ``look like'' either as the one of a wedge or as the one of the Minkowski space. It would be very interesting and important to understand whether this is due to the masslessness of the field or the result (i.e. the long range effect of the boundaries on the TT theory) is more deep.

This brings the importance of the explicit construction of the TT theory for a more general example - a massive free field. In the earlier version of \cite{Longo:2020amm} such attempt was made, though, unfortunately, later it was shown that the construction was flawed. So, currently there is no explicit answer for the massive modular Hamiltonian for the diamond region. This example has the crucial importance for many reasons. In relation to our problem, we can name at least a couple.

First of all, it is well-known, see e.g. \cite{Borchers:2000pv}, that in the massive case the Tomita-Takesaki theory will not have a geometric realization. In this case our definition cannot be directly adopted to define local temperature. Still, we believe that due to the locality of our construction, the approximate preferred observers could be defined allowing to extend our consideration to the massive case too. We expect that, at least formally, this definition will be closely related to the one considered in \cite{Arias:2016nip}.

Secondly, as we mentioned above, it would be very interesting to study the limits considered in the present paper in the massive case. This should help to understand whether the diamond boundaries continue to have the same drastic effect deep into the diamond even in the massive case. Physically, one can try to argue that because in the presence of mass, $m$, there is a natural scale, $1/m$, one should not ``feel'' boundaries for regions localized inside a diamond at the distances greater that this characteristic scale. But this naive consideration could be completely misleading, as our study of the massless case teaches.

Though some of the points raised about require the explicit knowledge of the Tomita-Takesaki theory, the others could be studied from a more general point of view. This is subject of our current studies and we hope to report on our progress in the future.

\section*{Acknowledgement}
AP acknowledges the partial support of CNPq under the grant no.312842/2021-0. The research of RM is supported by the CAPES doctorate fellowship. The authors have also benefited from discussions with Carolina Gregory and Cristian Landri.

\appendix

\bibliographystyle{utphys}
\bibliography{Ref}

\end{document}